\documentclass[proceedings, preprint]{rmaa}

\usepackage{paralist}

\usepackage{psfrag,color}

\newcommand{\hi}{\ion{H}{I}}

\newcommand{\hei}{\ion{He}{I}}
\newcommand{\heii}{\ion{He}{II}}
\newcommand{\oi}{\ion{O}{I}}
\newcommand{\oii}{\ion{O}{II}}
\newcommand{\oiii}{\ion{O}{III}}
\newcommand{\cii}{\ion{C}{II}}
\newcommand{\ciii}{\ion{C}{III}}
\newcommand{\civ}{\ion{C}{IV}}
\newcommand{\foiii}{[\ion{O}{III}]}
\newcommand{\foii}{[\ion{O}{II}]}
\newcommand{\fnii}{[\ion{N}{II}]}
\newcommand{\fsii}{[\ion{S}{II}]}
\newcommand{\fcliii}{[\ion{Cl}{III}]}
\newcommand{\neii}{\ion{Ne}{II}}
\newcommand{\fariv}{[\ion{Ar}{IV}]}

\SetYear{2010}
\SetConfTitle{XIII Latin American Regional IAU Meeting}

\title{Faint emission lines in planetary nebulae with a [WC] nucleus\altaffilmark{1}} 

\author{
  J. Garc\'{\i}a-Rojas,\altaffilmark{2} 
  M. Pe\~na,\altaffilmark{3}
  and M.~T. Ruiz\altaffilmark{4}}

\altaffiltext{1} {Based on data collected at Las Campanas Observatory, Carnegie Institution}
\altaffiltext{2}{Instituto de Astrof\'{\i}sica de Canarias, E-38200, La Laguna, Tenerife, Spain (jogarcia@iac.es).}
\altaffiltext{3}{Instituto de Astronom\'{\i}a, UNAM, Apdo. Postal 70-264, 04510, M\'exico D.F., M\' exico (miriam@astroscu.unam.mx).}
\altaffiltext{4}{Departamento de Astronom\'{\i}a, Universidad de Chile, Casilla Postal 36D, Santiago de Chile, Chile (mtruiz@das.uchile.cl).}

\shortauthor{Garc\'{\i}a-Rojas, Pe\~na, \& Ruiz}
\shorttitle{Faint emission lines in PNe with [WC] nucleus}

\listofauthors{Garc\'{\i}a-Rojas, Pe\~na, \& Ruiz}
\indexauthor{Garc\'{\i}a-Rojas}
\indexauthor{Pe\~na}
\indexauthor{Ruiz}

\abstract{We present first results from the analysis of a sample of 14 planetary nebulae with [WC] nucleus with detected faint carbon and oxygen recombination lines (RLs). The results are based on deep echelle spectra obtained with MIKE on the 6.5 m Magellan-Clay telescope in Chile. }

\resumen{Presentamos los primeros resultados del an\'alisis de una muestra de 14 nebulosas planetarias con n\'ucleo [WC] con l\'{\i}neas de recombinaci\'on d\'ebiles de ox\'{\i}geno y carbono. Los resultados est\'an basados en espectros profundos obtenidos con el espectr\'ografo MIKE en el telescopio Magallanes-Clay de 6.5 m en Chile.}

\addkeyword{ISM: abundances}
\addkeyword{planetary nebulae: general}
\addkeyword{stars: Wolf-Rayet}

\begin{document}
\maketitle

\section{Introduction}
\label{sec:intro}

The abundance discrepancy problem is a well known topic in modern astrophysics. Since almost 20 years ago, 
multiple studies, both in \ion{H}{II} regions and planetary nebulae (PNe) have found that ionic abundances of heavy element, such as C and O, obtained from the temperature and density independent ratios of optical recombination lines (RLs) are larger than that obtained following the canonical method, i.e. through the intensity of the strong collisionaly excited lines (CELs), which have a strong dependence with the temperature and in some cases, with density. During last years, several papers have debated the possible origins of this discrepancy, without reaching any certain conclusion \citep[see e.g.][and references therein]{garciarojasesteban07}. In PNe, one of the claimed explanations is the presence of H-deficient gas inclusions not fully mixed with the ambient gas of ``normal'' composition \citep{liuetal00}.

Only a few percentage ($\sim$15\%) of the known planetary nebulae (PNe) with studied stellar continuum are ionized by central stars showing Wolf-Rayet features. These stars are H-deficient and they show spectra with 
wide emission lines of C, O and He due to a strong wind characterized by a high mass-loss 
rate. They have been designed as [WC] stars.
This project is aimed to detect and measure faint RLs of C and O in PNe with central [WC] stars to search for any systematic difference between this type of objects with respect to ``normal'' PNe. One of the advantages of using heavy element RLs to compute the abundances is that we can derive the C/O ratio for all the objects in our sample. This ratio is usually difficult to derive because there are not carbon CELs in the optical range, and one has to search for UV spectroscopic data in the literature to derive the C abundance. Moreover, an additional objective of this project is to investigate if late evolution events (late or very late helium flashes) in the AGB phase of this stars could be the mechanism that injects the H-deficient gas necessary to explain the ADF in the planetary nebula.
 
\section{Observations}
\label{sec:observ}

High spectral resolution data were obtained at Las Campanas
Observatory (Carnegie Institution) with the 6.5-m telescope
Clay and the double echelle spectrograph Magellan Inamori
Kyocera Echelle (MIKE) on May, 2006, September 2009 and June 2010.
Details on the spectrograph setup and data reduction process can be found in \citet{garciarojasetal09}.

\section{First results}
\label{sec:results}

We detected and measured between 120 and 350 lines per object ---more than 3350 emission lines in total.  
Most of them correspond to usual CELs and RLs of  {\hi}, {\hei} and {\heii}, but we also detected a considerable amount of 
heavy element RLs, such as {\oi}, {\oii}, {\oiii}, {\cii}, {\ciii}, {\civ} and {\neii}.   
Line fluxes were measured by integrating all the flux between two given limits  
over a local continuum. In some blends, we applied multiple Gaussian profile fits, and in some particular cases, lines 
were so severely blended that we adopted the flux as the sum of all the lines.  
All the measurements were made using the SPLOT routine of IRAF
{\footnote{IRAF is distributed by the NOAO, which is operated by AURA, Inc., under contract to the National Science Foundation.}. 
More details of the analysis process are explained in \citet{garciarojasetal09}, where we have published the
complete study for the objects PB~8 and NGC~2867. 

We computed physical conditions, $T_{\rm e}$ and $n_{\rm e}$ using several diagnostic ratios. In general, we adopted 
$T_{\rm e}$({\fnii}) for the low ionization zone and $T_{\rm e}$({\foiii})  for the medium-high 
ionization zone. In the most ionized objects, we also used $T_{\rm e}$({\fariv}) for the highest ionization zone. 
Regarding the density, we have obtained several values from different diagnostic ratios, i.e., {\fsii} and {\foii} ratios for the low 
ionization zone,  {\fcliii} ratio for the medium-high ionization zone and {\fariv} ratio for the highest ionization ions. 
Taking into account that the first two ratios saturate at densities $>$10$^4$ cm$^{-3}$,  we adopted $n_{\rm e}$({\fcliii}) as more representative of the low to intermediate ionization zones of  nebulae for those PNe with higher densities (see below). We then computed ionic abundances for a large number of ions from CELs and total abundances by using ionization correction factors from the literature. 

\begin{figure}[!t]
  \includegraphics[width=\columnwidth]{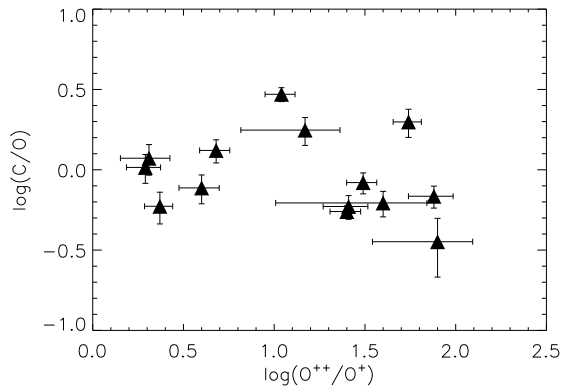}
  \caption{C/O ratio $vs.$ ionization degree.}
  \label{fig1}
\end{figure}

We derived C$^{++}$/H$^+$ and O$^{++}$/H$^+$ ratios from pure {\cii} and {\oii} RLs for all the objects in the sample. Moreover, for some objects we could also derive C$^{+3}$/H$^+$ ratios from RLs. The very high spectral resolution and depth of our spectra allowed us to measure with a proper deblending of telluric lines, the very faint {\oi} multiplet 1 RLs and hence, to derive the O$^+$/H$^+$ ratios for 7 objects using pure RLs. This is the largest sample of objects with derived O$^+$ abundance from RLs in the literature. 
With all these data, we have been able to compute C/O ratios for our sample from optical RLs. In Fig.~\ref{fig1} we can see that there are 6 objects with C/O $>$ 1, which indicates that about 43\% of the sample are C-rich. This fraction is very similar to the one derived from the {\ciii}] $\lambda$ 1909)/{\foiii} $\lambda$5007 line ratios for a large sample of PNe by \citet{rolastasinska94}.  

We computed the abundance discrepancy factor \citep[ADF, following the definition in][]{liuetal00} for O$^{++}$ for all our objects. 
The values obtained for the ADF(O$^{++}$) range from 1.4 to 3.9 and are moderate and similar to the typical value obtained in PNe. 

We could also derive the ADF(O$^+$) for seven objects of our sample. Given the low critical densities of the {\fsii} and {\foii} lines used to compute densities, we have to manage with caution with the densities obtained from these ratios, especially when computing abundances from density-sensitive collisionally excited lines, such as {\foii} $\lambda$$\lambda$3726+29. In Fig.~\ref{fig2} we show the dependence of the ADF with the assumed density for the O$^+$ zone. This strong dependence is due to O$^+$/H$^+$ computed from {\foii} CELs is extremely dependent of the assumed density, whether the same abundance ratio obtained from pure RLs is almost independent on the electron density.

\begin{figure}[!t]
  \includegraphics[width=\columnwidth]{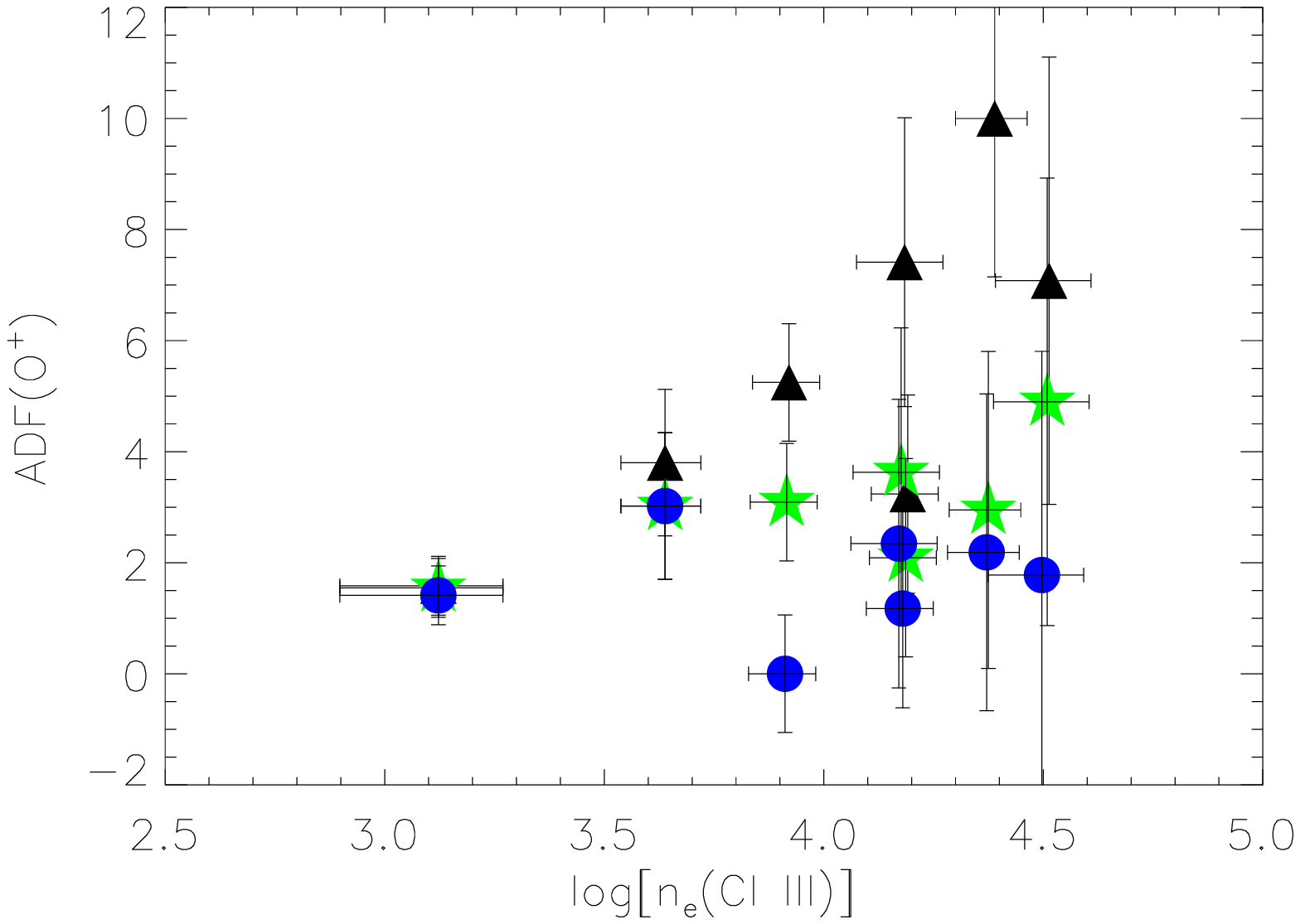}
  \caption{ADF(O$^+$) computed by assuming $n_{\rm e}$({\foii}) (black triangles), $n_{\rm e}$({\fsii}) (green stars) or $n_{\rm e}$({\fcliii}) (blue circles.}
  \label{fig2}
\end{figure}

\end{document}